\documentclass[aps,prd,twocolumn,showpacs,showkeys,preprintnumbers,amsmath,amssymb,groupedaddress]{revtex4}

\usepackage[dvips]{graphicx}
\usepackage{dcolumn}
\usepackage{bm}

\def\Journal#1#2#3#4{{\em #1} {\bf #2}, #3 (#4).}

\begin{document}

\title{Can lightning be a noise source \\ for a
 spherical gravitational wave antenna? }

\author{Nadja Sim\~ao Magalh\~aes}

\email{nadjam@ig.com.br}

\affiliation{Centro Federal de Educa\c{c}\~ao Tecn\'{o}logica de
S\~ao Paulo, Rua Pedro Vicente 625, S\~ao Paulo, SP 01109-010,
Brazil and \\
Instituto Tecnol\'ogico de Aeron\'autica - Departamento de
F\'{\i}sica\\ P\c ca. Mal. Eduardo Gomes 50, S\~ao Jos\'e dos
Campos, SP 12228-900, Brazil}

\author{Rubens de Melo Marinho, Jr. }

\affiliation{Instituto Tecnol\'ogico de Aeron\'autica -
Departamento de F\'{\i}sica\\ P\c ca. Mal. Eduardo Gomes 50, S\~ao
Jos\'e dos Campos, SP 12228-900, Brazil}

\author{Odylio Denys de Aguiar}

\affiliation{Instituto Nacional de Pesquisas Espaciais - Divis\~ao
de Astrof\'{\i}sica\\Av. dos Astronautas 1758, S\~ao Jos\'e dos
Campos, SP 12227-010, Brazil}

\author{Carlos Frajuca}

\affiliation{Centro Federal de Educa\c{c}\~ao Tecn\'{o}logica de
S\~ao Paulo, Rua Pedro Vicente 625, S\~ao Paulo, SP 01109-010,
Brazil}


\begin{abstract}
The detection of gravitational waves is a very active research
field at the moment. In Brazil the gravitational wave detector is
called Mario SCHENBERG. Due to its high sensitivity it is
necessary to model mathematically all known noise sources so that
digital filters can be developed that maximize the signal-to-noise
ratio. One of the noise sources that must be considered are the
disturbances caused by electromagnetic pulses due to lightning
close to the experiment. Such disturbances may influence the
vibrations of the antenna's normal modes and mask possible
gravitational wave signals. In this work we model the interaction
between lightning and SCHENBERG antenna and calculate the
intensity of the noise due to a close lightning stroke in the
detected signal. We find that the noise generated does not disturb
the experiment significantly.
\end{abstract}

\pacs{04.80.Nn,42.20.Jb,42.25.Bs,92.60.Pw,95.55.Ym}

\keywords{Gravitational wave detectors; electromagnetic wave
propagation, transmission and absorption; atmospheric
electricity.}

\maketitle

\section{SCHENBERG's features}

The detection of gravitational waves is presently an active
research field and there are several detectors either being built
or in operation around the world, involving countries like USA,
Japan, Germany, The Netherlands, Italy and Australia. In Brazil
the gravitational wave detector is called Mario SCHENBERG. This
resonant-mass detector will operate in coincidence with some of
the detectors in other countries. It is now in an advanced process
of installation at the University of Sao Paulo, in the capital of
Sao Paulo State. It is expected that SCHENBERG will be tested with
three of its nine transducers in 2006.

Resonant-mass gravitational wave detectors like SCHENBERG detect
the presence of a signal basically through the vibration of its
first quadrupolar modes. Transducers strategically positioned on
the antenna's surface are able to transform these vibrations into
electric signals that can be digitally processed. As one can see
from Figure \ref{fig:schen}, SCHENBERG's antenna is spherical; it
is made of a CuAl alloy (6\% Al) and has 65 cm in diameter. More
details on its functioning can be found in reference
\cite{ody2004}.

According to General Relativity the quadrupolar modes are expected
to be excited by gravitational waves \cite{mis}. Due to its
characteristics in SCHENBERG the first of these modes are tuned to
$ \approx 3.2 $ kHz. However, if an electromagnetic pulse
generated by a lightning stroke hits the antenna most of its
normal modes should be excited, including the quadrupolar ones.
Therefore lightning is expected to create noise in the detector
perhaps masking detectable gravitational waves.

In a previous work we have modelled the interaction between cosmic
rays and SCHENBERG \cite{marinho2001}. In that paper the
interaction was basically thermodynamical. In this work the
approach is different: the electromagnetic wave generated by the
lightning stroke hits the antenna and transfers momentum to it. As
a first approximation to the calculations we will assume that
SCHENBERG consists only of the spherical antenna, ignoring for now
the  cryogenic chambers.

In what follows, after a brief description of how SCHENBERG would
respond to the influence of an electromagnetic wave we will
present the mathematical model for lightning strokes. Then the
lightning force density is calculated since it will be used in the
model of the interaction between the detector and the lightning.
Results and conclusions are the closing sections.

\section{Electromagnetic waves in conducting media}

A lightning creates an electromagnetic pulse that propagates
 with the speed of light. We want to investigate what
happens when this pulse reaches the antenna, which is a conducting
solid medium. For the description of the behavior of
electromagnetic waves in conducting media one can use  results
found in the literature \cite{jack}. Two types of losses are
encountered by an electromagnetic wave striking a metallic
surface: 1) the wave is partially reflected from the surface
(reflection loss), and 2) the transmitted portion is attenuated as
it passes through the medium. This latter effect is called
absorption loss. Absorption loss is the same in either the near or
the far field and for electric or magnetic fields. Reflection
loss, on the other hand, is dependent on the type of field and the
wave impedance.

The transition region between the near and the far fields is
around $\lambda / 2 \pi$, where $\lambda$ is the source's
wavelength. In the case of SCHENBERG, the relevant wavelength is $
\lambda  = {c \mathord{\left/
 {\vphantom {c \nu _0}} \right.
 \kern-\nulldelimiterspace} \nu _0} = {{3 \times 10^8 } \mathord{\left/
 {\vphantom {{3 \times 10^8 } {3175}}} \right.
 \kern-\nulldelimiterspace} {3175}} \simeq 9.4 \times 10^4$ m,
 thus implying a transition region around 15 km.
We will consider distances smaller than 15 km because they include
stronger strokes. Within this distance the far field limit would
be suited only for higher frequencies, which are  much weaker then
the lower ones, as one can see from Figure \ref{fig:wpoisson}. For
these reasons we conclude that the electromagnetic fields
generated by  lightning that hit the ground close to the detector
are in the near field limit.

\subsection{Absorption loss}

It can be shown that when an electromagnetic wave reaches a good
conductor it is reduced to $1/e\, = \, 0.369$ of its initial
amplitude after travelling the distance (all units in this paper
are in the MKS system)
\[
\delta  \simeq \sqrt {\frac{2} {{\mu \omega \sigma }}}.
\]
The distance $\delta$ is called {\it skin depth}. It varies with
the frequency of the wave, as can be seen in Table
\ref{table:skin} \cite{ott}. The symbols $\mu$ and $\sigma$ refer,
respectively to the permeability (equal to $4 \pi \times 10^{-7}$
H/m for free space) and to the conductivity (equal to $5.82 \times
10^7$ mhos/m for copper).

Then for a medium with thickness $t$ the absorption loss is given
by
\[
A = 8.69\left( {\frac{t} {\delta }} \right)\quad dB.
\]
In the case of SCHENBERG $\delta \sim 1.5 $ mm. Thus an
appreciable fraction of absorption loss should happen practically
at  the antenna's surface.

\subsection{Reflection loss in the near field}

In the near field the electric ($E$) and the magnetic ($H$) fields
must be considered separately, since the ratio of the two is not
constant. In this limit reflection loss varies with the wave
impedance, defined by $Z \equiv E/H$. For the same source
frequency a high-impedance (electric) field has a higher
reflection loss than a low-impedance (magnetic) field.

In the case of SCHENBERG, mostly made of copper and sensitive to
frequencies around 3.2 kHz, we can expect a minimum reflection
loss around 50 dB \cite{ott}. In fact, for low frequency plane
waves, as the ones considered here, reflection loss accounts for
most of the attenuation.

\section{Model for the lightning}

The phenomenon of lightning is still not fully understood but
several models have already been proposed to explain it
\cite{lin80}. In brief, a cloud-to-ground lightning flash is
usually composed of several intermittent discharges called
strokes, which are made up of a leader phase and a return stroke
phase. The leader initiates the return stroke, which is an upward
travelling wave. A relatively large electric field exists in the
leader-return stroke channel and this field produces ionization
and results in a current. The power input renders the channel very
luminous and causes its rapid expansion, producing thunder. More
details on the phenomenon can be found in \cite{uman69}.

In our analysis we will consider the effect of a first return
stroke on SCHENBERG. The idealized stroke is drawn in Figure
\ref{fig:stroke}, adapted from \cite{lin80}.

The electric field intensity ${\mathbf E}$ and the magnetic flux
intensity ${\mathbf B}$ produced by the channel current $i(z,t)$
at a distance $D$ from the vertical channel of height $H$ are
given by \cite{uman75}

\[
\mathbf E(D,t) = \frac{1} {{2\pi \epsilon _0 }}\left[
{\int\limits_0^H {\int\limits_0^t {\frac{{2 - 3\sin ^2 \theta }}
{{\rho ^3 }}} i(z,\tau  - \rho /c)d\tau \,dz} } \right.
\]
\[
 + \int\limits_0^H {\frac{{2 - 3\sin ^2 \theta }}
{{c\rho ^2 }}} i(z,t - \rho /c)\,dz
\]
\begin{equation}
\left. { - \int\limits_0^H {\frac{{\sin \theta }} {{c^2 \rho
}}\frac{{\partial i(z,t - \rho /c)}} {{\partial t}}} \,dz} \right]
\mathbf{e}_z ,
 \label{eq:E}
\end{equation}

\[
\mathbf B(D,t) = \frac{{\mu _0 }} {{2\pi }}\left[ {\int\limits_0^H
{\frac{{\sin \theta }} {{\rho ^2 }}} } \right.i(z,t - \rho /c)\,dz
\]
\begin{equation}
\left. { - \int\limits_0^H {\frac{{\sin \theta }} {{c\rho
}}\frac{{\partial i(z,t - \rho /c)}} {{\partial t}}} \,dz} \right]
\mathbf{e}_\phi, \label{eq:B}
\end{equation}
where $\epsilon _0$ is the permittivity, $\mu _0$ the permeability
of free space and $c$ is the speed of light in vacuum.

There are several models for the channel current $i(z,t)$
\cite{lin80}, all trying to describe actual strokes as accurately
as possible. Since we are interested mostly in the intensity of
the disturbance caused by a stroke in the detection of
gravitational wave process we will use a simple expression for the
current, which accounts for the basic features of a typical
lightning.

 We assume that the stroke has a typical length
of 5 km. It takes $10 \mu$s to reach the maximum current of 30 kA
and lasts typically $50 \mu$s. With these values we model the
current as a function of time as a Poisson distribution:
\begin{equation} \label{eq:poisson}
i(t) = a\,t\,e^{ - \lambda t},
\end{equation}
with $a=8.155\times 10^9$A/s and $\lambda = 10^5$ s$^{-1}$. See
Figure \ref{fig:poisson} for the plot of this function. We will
adopt here the Bruce-Golde model \cite{brucegolde}, for which
$i(z,t)=i(0,t)$. Also, we will assume that the lightning stroke is
quite close to the detector, at a distance $D= 1500$ m, which
ensures a high amplitude to the electromagnetic front that will
disturb the antenna.

\section{Determination of the lightning's force density  }

From the expressions (\ref{eq:E}), (\ref{eq:B}) and
(\ref{eq:poisson}) we were able to compute the Poynting vector for
the electromagnetic (e-m) radiation that reaches SCHENBERG using
the formula $ \mathbf S(t) = \mu _0^{-1} \mathbf E(t) \times
\mathbf B(t) $. The coordinate system used is depicted in Figure
\ref{fig:stroke}, with $D$ fixed at $1.5$ km. From this expression
we found the density of momentum of the wave, $\mathbf p(t) =
\mathbf S(t)/c^2$. The plots of the electric and magnetic fields
at this close distance are shown in Figure \ref{fig:eb}.

By differentiating $\mathbf p(t)$ we obtained the expression for
force density of the e-m wave, $\mathbf f(t)$, which for the
distance we chose has modulus
\[
f(t) = 2.21 \times 10^{ - 3} e^{ - 10^{ - 5} t} \left( {t - 10^{ -
5} } \right) +
\]
\begin{equation}
 + e^{2 \times 10^{ - 5} t} \left( { - 2.83 \times 10^3 \;t^2
 + 0.768\;t - 2.42 \times 10^{ - 7} } \right)
\end{equation}
with units of $V T s^2 m^{-3}$ in the MKSA system.

\section{The model of the interaction}

The mechanical model of a spherical gravitational wave antenna has
already been presented in the literature. We will follow here the
notation on \cite{magalhaes97} for the case that there are no
transducers coupled to the antenna and that only source of force
on the sphere is the e-m wave. Then the Fourier transform of the
radial amplitude of motion of the detector surface under the
influence of this wave is given by
\begin{equation}
\tilde a_m (\omega ) = k{\kern 1pt} I_m \left( { - \omega ^2  +
i\frac{\omega _0}{Q} \omega  + \omega _0^2 } \right)^{ - 1} \tilde
f(\omega ), \label{eq:aw}
\end{equation}
where $\tilde f(\omega )$ is the Fourier transform of the force
density, the constant $k$ is defined by
\[
k \equiv \alpha (R){\kern 1pt} R^2 /\left( {\varrho M_{eff} N}
\right)
\]
and
\[
I_m  = \int\limits_{\theta  = 0}^\pi  {\int\limits_{\phi  = 0}^\pi
{Y_{2m} \left( {\theta ,\phi } \right)} } \sin ^2 \theta \;\sin
\phi \;d\theta \;d\phi.
\]
The list of constants used in the equations is presented in Table
\ref{table:const}. They are based on recent data from SCHENBERG
\cite{tesecesar}. The functions $Y_{2m} \left( {\theta ,\phi }
\right)$ are the spherical harmonics and can be found in textbooks
\cite{jack} . As usual, $\omega _0 = 2 \pi \nu _0$. The limits of
integration are such that they cover the half-sphere that is in
the direction of the lightning stroke.

\section{Results}

From equation (\ref{eq:aw}) we could determine the amplitude of
the motion caused by the e-m wave ($a_m(t)$) in the five
degenerate quadrupole modes of SCHENBERG, given by the index $m =
-2,-1,0,1,2$. We found that modes m=1 and m=-1 are not disturbed
at all by this wave. Modes m=2 and m=-2 are equally disturbed,
slightly more than mode m=0.

In gravitational wave detection the adimensional amplitude
$h_m=a_m/R$ is preferred as a measure of the antenna's motion. In
terms of this figure the results we obtained on the motion of the
modes with time is shown in Figure \ref{fig:h_m}. Notice that the
modes present a maximum adimensional amplitude around $3.4 \times
10^{-24}$.

\section{Conclusions}

The gravitational wave detector SCHENBERG is expected to reach a
maximum sensitivity of $h \sim 10^{-20}$ for impulsive waves when
fully operational in a few years. The noise caused by a typical,
close lightning stroke on the detector was found in this work to
be approximately three orders of magnitude smaller than this
expected sensitivity.

Had the electromagnetic shielding due to the metallic cryogenic
chambers  be taken into account in the calculations this noise
would be even less significant. Therefore we conclude that it is
unlikely that lightning strokes should cause detectable noise
while SCHENBERG is running.

This result is relieving since it may happen that SCHENBERG does
not run in coincidence with other detectors from time to time, and
it must be free of as much kinds of noise as possible. The region
in which this detector is been built is prone to summer
thunderstorms during approximately 2 months, accompanied by many
lightning strokes. We have just shown that such strokes should not
directly disturb the data significantly. Occasionally electric
fluctuations may happen in power supply due to lighting strokes
but these can be easily ruled out with the continuous monitoring
of the experiment.

Although the maximum adimensional amplitude due to lighting
strokes found above is larger than the amplitudes due to many
monochromatic gravitational waves \cite{mis} this kind of noise
should not disturb the detection of this kind of waves since the
signal is integrated in time.

\begin{acknowledgments}
ODA acknowledges FAPESP and CNPq for financial support (grants \#
1998/13468-9 and 306467/2003-8, respectively) . The authors are
thankful to Osmar Pinto, Jr., and Fernando J. Miranda for the
references on lighting stroke models and information on typical
values that characterize lightning.
\end{acknowledgments}


\newpage

\begin{thebibliography}{9}


\bibitem{ody2004}{O. D. Aguiar et al., \Journal{Class. Quantum Grav.}{21}{S457}{2004}}

\bibitem{mis}{C. W. Misner, K. S. Thorne and J. A. Wheeler. {\it Gravitation}, New York: Freeman, 1973.}

\bibitem{marinho2001}{R. M. Marinho, Jr, N. S. Magalhaes, O. D. Aguiar and C. Frajuca, \Journal{Phys. Rev.}{D64}{065017}{2001}}

\bibitem{jack}{J. D. Jackson. {\it Classical Electrodynamics}, 2nd ed., New York: Wiley, 1975.}

\bibitem{ott}{H. W. Ott. {\it Noise reduction techniques in electronic systems}, New York: Wiley, 1976. Chapter 6.}

\bibitem{lin80}{Y. T. Lin, M. A. Uman and R. B. Standler, \Journal{J. Geophys. Res.}{85}{1571}{1980}}

\bibitem{uman69}{M. A. Uman. {\it Lightning}, New York: McGraw-Hill, 1969.}

\bibitem{uman75}{M. A. Uman, D. K McLain and E. P. Krider,
\Journal{Am. J. Phys.} {43}{33}{1975}}

\bibitem{brucegolde}{C. E. R. Bruce and R. H. Golde,
\Journal{J. Inst. Elec. Eng.} {88}{487}{1941}}

\bibitem{magalhaes97}{N. S. Magalhaes, O.D. Aguiar, W.W. Johnson and C. Frajuca,
\Journal{Gen. Relat. Grav.} {29}{1511}{1997}}

\bibitem{tesecesar}{C. A. Costa. {\it A mathematical model for the mechanical
behavior of MARIO SCHENBERG gravitational wave detector}, Sao Jose
dos Campos : INPE, 2002. (MsC. thesis)}.

\end{thebibliography}

\newpage
\begin{table*}[h]
\begin{tabular}{ccc}
 \hline
  Frequency & $\delta$  & $\delta$ \\
  (Hz) & for Cu & for Al \\
 \hline
  60  & 8.51 & 10.90 \\
  100  & 6.60 & 8.46 \\
1 k & 2.08 & 2.67 \\
 3.2 k & 1.63 & 1.49 \\
 10 k & 0.66 & 0.84 \\
 100 k & 0.20 & 0.28 \\
 1 M & 0.08 & 0.08 \\
 10 M & 0.02 & 0.02 \\
\hline
\end{tabular}
\caption{\label{table:skin} Skin depth $\delta$ for copper and
aluminum, in mm.}
\end{table*}

\begin{table*}
\begin{tabular}{ccc}
 \hline
 Name & Symbol  & Value \\

 \hline
 Radius at 4K  & R & 0.3239 m \\
  Geometric constant  & $\alpha (R)$ & 2.862 \\
Density at 4K & $\varrho$ & $8077.5 $ kg/m$^3$ \\
 Effective mass & $M_{eff}$ & 288 kg \\
Normalization constant & N & $0.142 $kg$^3$ \\
Quadrupole mode frequency & $\nu _0$ &  3175 Hz \\
Mechanical quality factor & Q & $2 \times 10^7$ \\

 \hline
\end{tabular}
\caption{\label{table:const} Values of the constants used in the
calculations.}
\end{table*}

\newpage

LIST OF FIGURE CAPTIONS:\\

\begin{figure}
    \begin{center}
        \includegraphics[width=8.5cm]{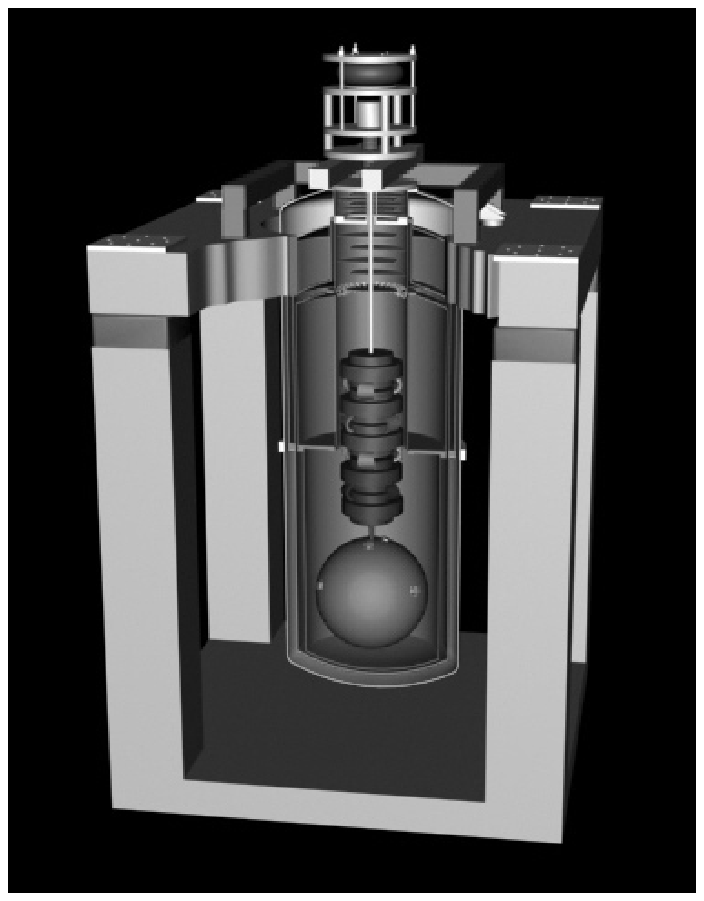}
        \caption{} \label{fig:schen}
    \end{center}
\end{figure}

Figure \ref{fig:schen}: Schematics of the SCHENBERG detector. The
spherical antenna is carefully suspended inside metallic chambers
that keep it in vacuum at 4 K.

\begin{figure}
    \begin{center}
        \includegraphics[width=8.5cm]{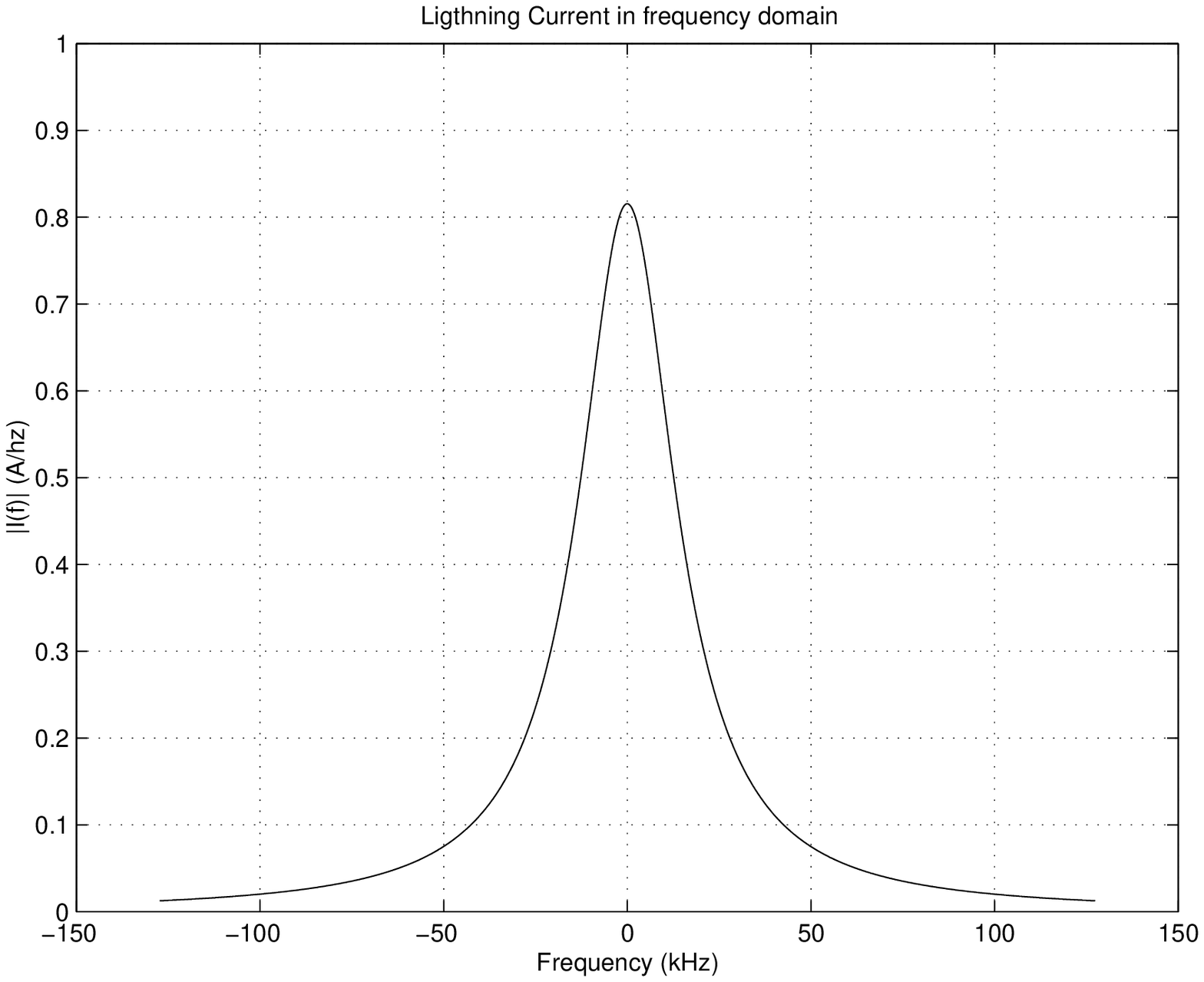}
        \caption{} \label{fig:wpoisson}
    \end{center}
\end{figure}

Figure \ref{fig:wpoisson}: Plot  of the Fourier transform of the
lightning's current.

\begin{figure}
    \begin{center}
        \includegraphics[width=8.5cm]{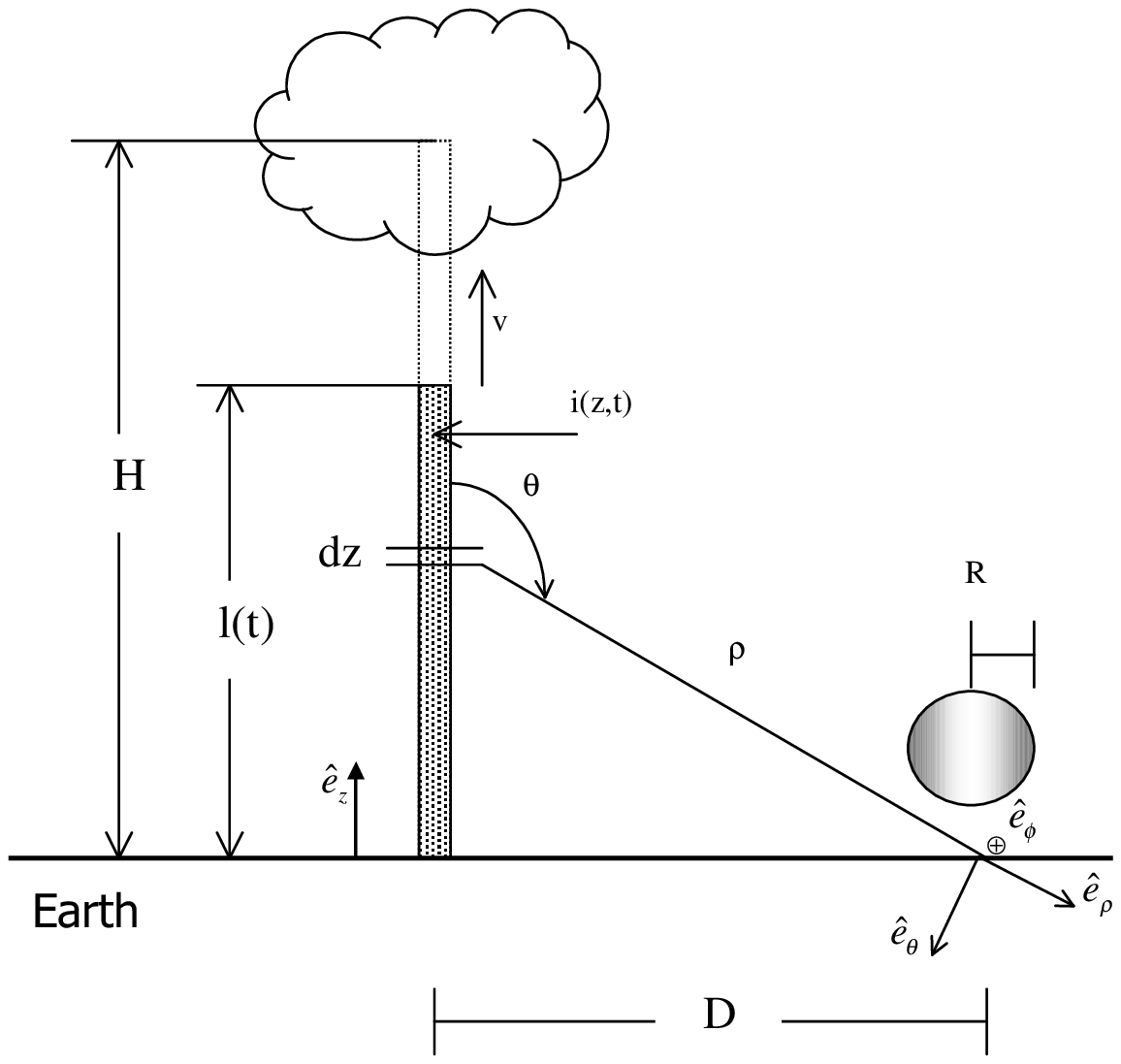}
        \caption{}
        \label{fig:stroke}
    \end{center}
\end{figure}

Figure \ref{fig:stroke}: Schematics of an idealized return stroke
with the geometrical and physical parameters used in the
calculations.

\begin{figure}
    \begin{center}
        \includegraphics[width=5.5cm]{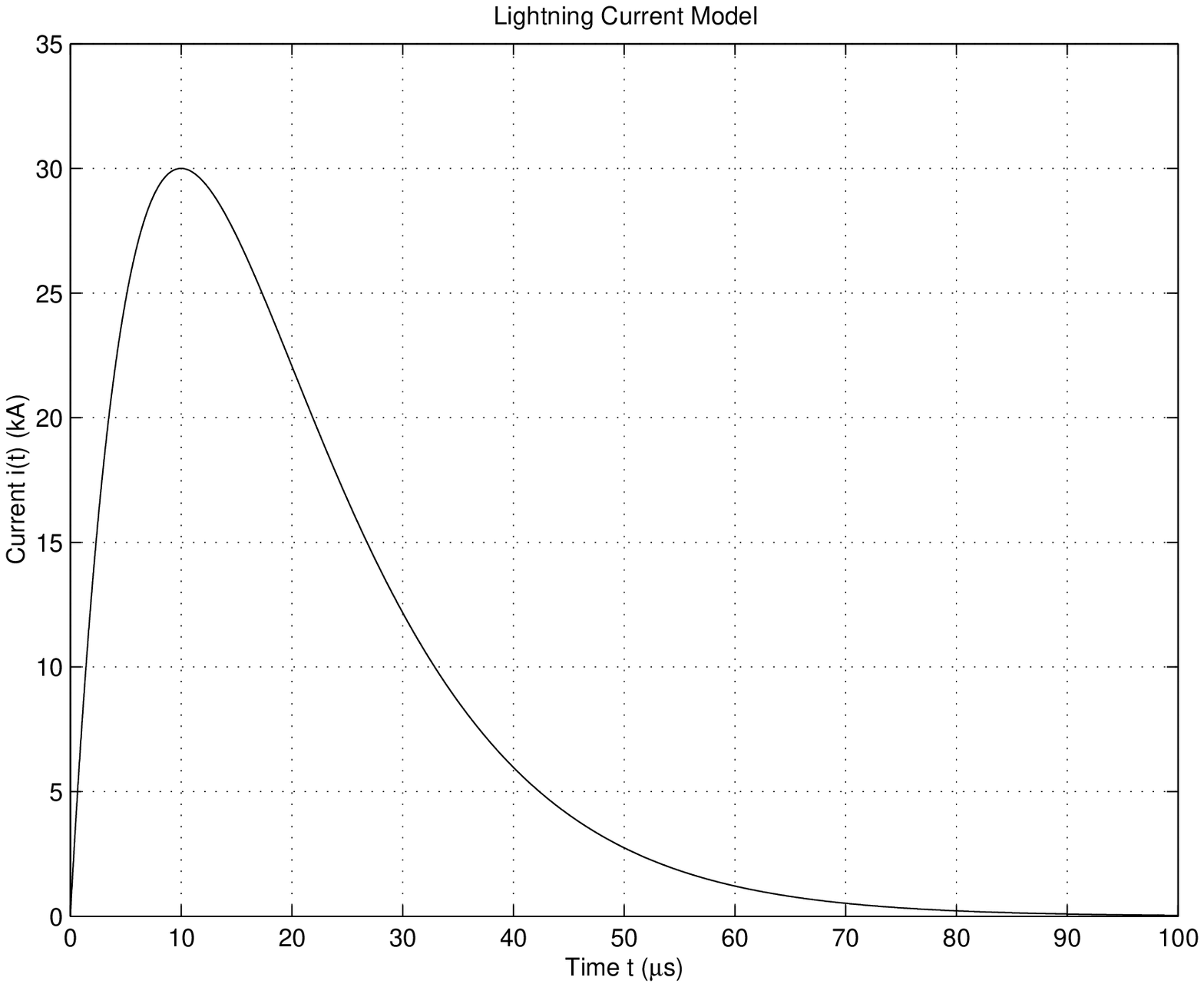}
        \caption{}
        \label{fig:poisson}
    \end{center}
\end{figure}

Figure \ref{fig:poisson}: Plot of the behavior of the current of
the lightning as a function of time, modelled according to a
Poisson distribution.

\begin{figure}
    \begin{center}
         \includegraphics[width=5.5cm]{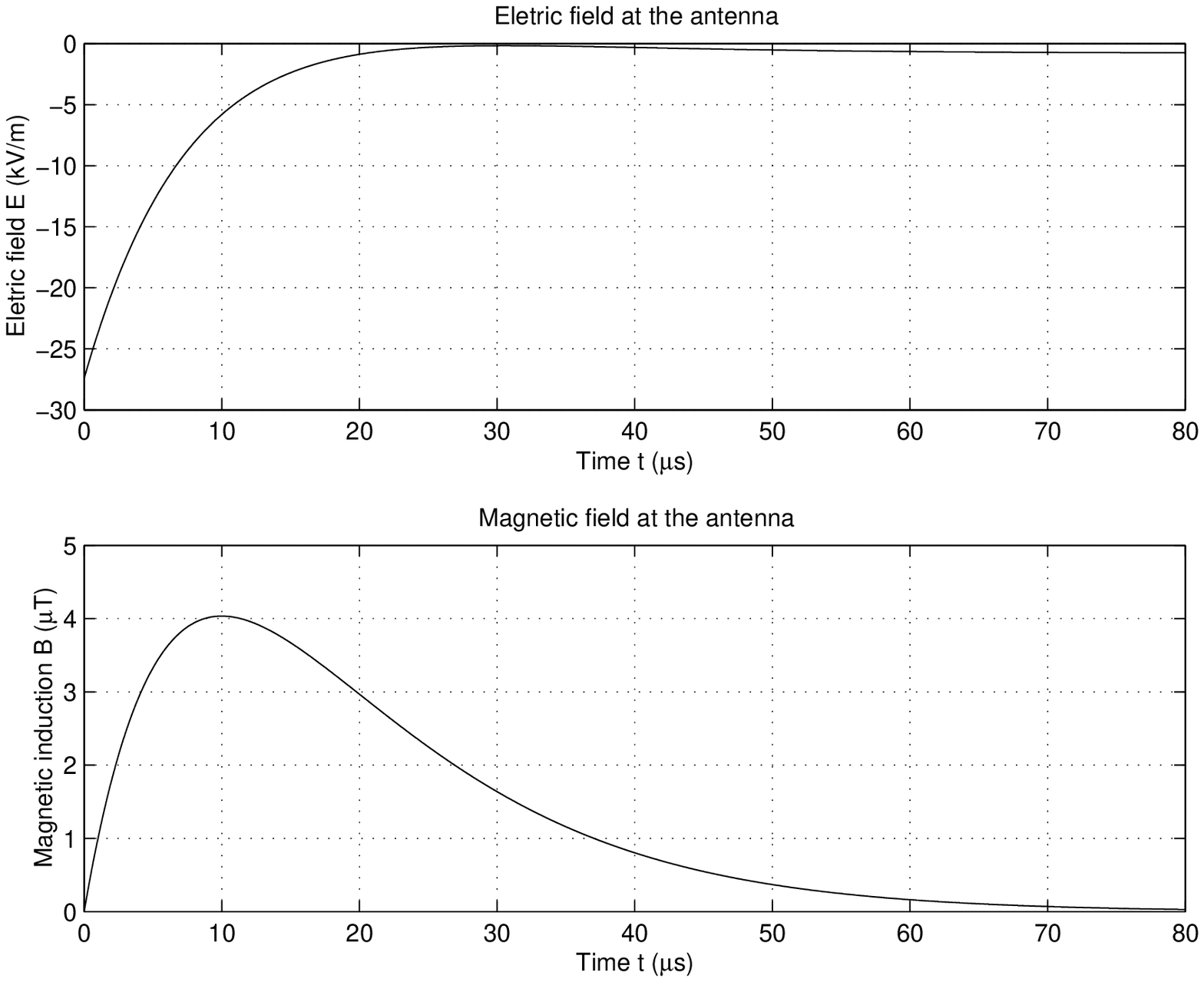}
         \caption{} \label{fig:eb}
     \end{center}
\end{figure}

Figure \ref{fig:eb}: Plots of the electric (a) and magnetic (b)
fields as functions of time for D = 1500 m.

\begin{figure}
    \begin{center}
        \includegraphics[width=5.5cm]{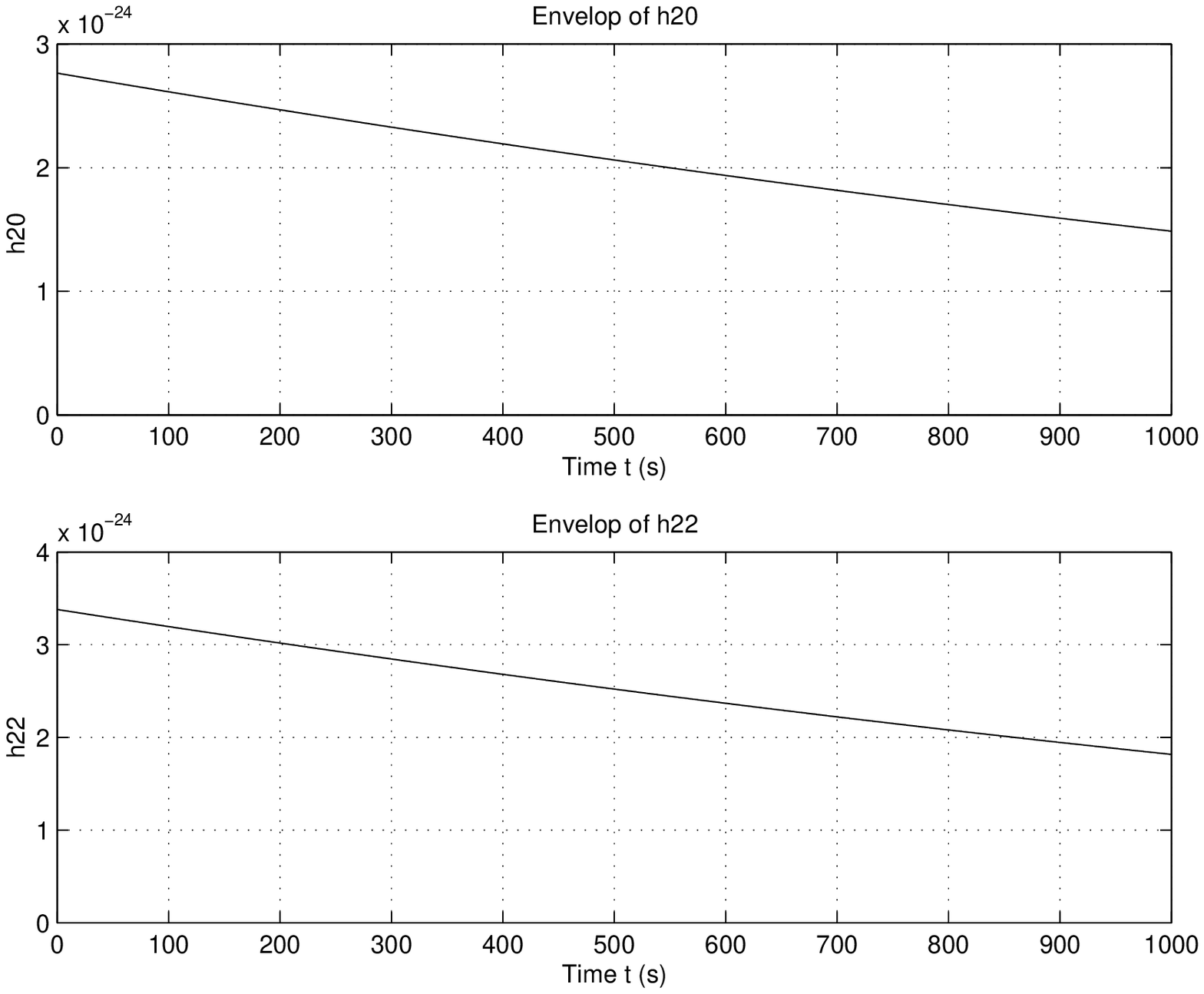}
        \caption{}
        \label{fig:h_m}
    \end{center}
\end{figure}

Figure \ref{fig:h_m}: Adimensional envelop of amplitudes as a
function of time for modes (a) m=0 and (b) m=2. Notice that the
amplitude decays slowly due to the antenna's high mechanical
quality factor.

\newpage

\end{document}